\documentclass{he_symp}
\usepackage{psfig,graphicx,epsfig}
\usepackage{color}
\usepackage{amsmath,amssymb,epic,eepic,array}
\begin{document}
\renewcommand{\FirstPageOfPaper }{ 91}\renewcommand{\LastPageOfPaper }{ 99}
\newcommand{\g}{$\gamma$\ }
\newcommand{\gr}{$\gamma$-ray\ }
\newcommand{\grs}{$\gamma$-rays\ }
\newcommand{\grpsr}{$\gamma$-ray pulsar\ }
\newcommand{\grpsrs}{$\gamma$-ray pulsars\ }
%
%
\title{Gamma-Ray Pulsars}
\author{Gottfried Kanbach }  
\institute{Max--Planck--Institut f\"ur extraterrestrische Physik, Giessenbachstra{\ss}e, 85741 Garching, Germany
}
\maketitle

\begin{abstract}

Gamma-ray photons from young pulsars allow the deepest insight into the properties and interactions of high-energy 
particles with magnetic and photon fields in a pulsar magnetosphere. Measurements with the Compton Gamma-Ray 
Observatory have led to the detection of nearly ten \grpsrs . Although quite a variety of individual signatures is 
found for these pulsars, some general characteristics can be summarized: (1) the \gr lightcurves of most high-energy 
pulsars show two major peaks with the pulsed emission covering more than 50\% of the rotation, i.e. a wide beam of 
emission; (2) the \gr spectra of pulsars are hard (power law index less than 2), often with a luminosity maximum 
around 1 GeV. A spectral cutoff above several GeV is found; (3) the spectra vary with rotational phase indicating 
different sites of emission; and (4) the $\gamma$-luminosity scales with the particle flux from the open regions of the 
magnetosphere (Goldreich-Julian current).      

\end{abstract}

\section{Introduction}

If the perceptiveness of Jocelyn Bell had not led to the discovery of radio pulsars in 1967, the three brightest 
localized \gr sources would have certainly attracted the attention of the astronomical community. The first 
high-energy \gr maps of the Galaxy (SAS-2, Fichtel et al.,1975; COS-B, Mayer-Hasselwander et al., 1982) are 
dominated by the sources now known as the young pulsars Crab, Vela and Geminga. Unlike in any other waveband these 
young pulsars outshine other sources at energies above 100 MeV. The sky survey performed with the Compton Gamma-Ray 
Observatory (CGRO) from 1991 to 2000 led to the detection of at least 4 more gamma-ray pulsars, raising the total 
number to 7 sources. Listing also some detections of lower significance we now have about a dozen potential \grpsrs 
. This is still a small number compared to the known population of radio ($\sim $1500) or X-ray ($\sim $30) pulsars. 
However the $\gamma$-radiation should provide the deepest insights into the physics of the extreme astrophysical site 
characterized by magnetic fields close to the quantum-mechanical scale ($B_{cr} = {m^2 c^3} / {e \hbar } = 
4.414\times 10^{13}\  G$), space-time near to a fast rotating neutron star, and relativistic speeds in the outer 
co-rotating magnetosphere. As we will see, the number of detected \grpsrs is quite in agreement with expectations 
based on sensitivities of current telescopes. Moreover, it is quite likely that several of the unidentified \gr  
sources, which belong mostly to a galactic population and are strongly correlated with population II regions (i.e. 
molecular clouds, OB associations, SNRs), are \grpsrs  without radio emission.

\section{High Energy Pulsar Physics} 

Why can, indeed must, we expect high-energy emissions from young pulsars? 
Goldreich and Julian (1969) have shown that the magnetosphere of a pulsar is filled with a charge-separated plasma. 
A rotating, magnetized, conducting neutron star works as a unipolar inductor. The positive and negative charges 
inside the star, redistributed by the Lorentz forces, arrange  themselves such that the electric field from charge 
separation counterbalances the magnetic forces and no permanent currents flow inside the star. The interior charge 
distribution determines the electric fields outside the star, requiring continuity for the potentials and tangential 
electric fields at the stellar surface. The radial components of this magnetospheric electric field ($\sim 6 \times 
10^{10} P_s^{-1}$ V cm$^{-1}$) have a discontinuity on the stellar surface which implies a surface charge layer. The 
electric fields close to the surface greatly surpass the gravitational attraction as well as the exit work functions 
of charges from the star. Therefore the magnetosphere of the neutron star is  filled with a charge separated plasma 
which tends to compensate the induced electric fields until $\textbf{E}\cdot \textbf{B}=0$ is achieved and the 
magnetosphere comes to equilibrium. The required charge density, called the Goldreich--Julian density, is given by 
\begin{equation}  n_{GJ} \simeq - {{\vec{\omega} \times \vec{B}} \over {2 \pi  c}} = 7\times 10^{10} B_{\| ,12} 
P_s^{-1}\  {\rm cm}^{-3}, \end{equation}
 where $B_{\| ,12}$  (units of $10^{12}$ G) is the field component parallel to $\vec{\omega}$ and $P_s$ is the 
rotation period in s.

However there can be no static equilibrium in the co-rotating magnetosphere of a pulsar. At a distance of $r_c = c / 
\omega \sim 5\times 10^4 P_s$ km from the axis of rotation (the \lq light cylinder\rq ), particles and fields would 
have to co-rotate with the speed of light. Relativistic effects of retardation and plasma mass loading will severely 
distort the pulsar magnetosphere close to the light cylinder and field lines trying to extend across $r_c$ are 
forced open to the outside. Regions of the magnetosphere threaded by these field lines (\lq open regions\rq) release 
their charges into the pulsar wind zone. The outflow leads to a deficit of charges in the regions above the magnetic 
poles (\lq polar cap\rq) and between the zero charge density surfaces ($n_{GJ} = 0$) and the equatorial 
magnetosphere close to $r_c$  (\lq outer gaps\rq ). In these gaps the electrostatic potential of the rotating dipole 
is not balanced by charges and is available to accelerate particles to very high energies. Additional accelerating 
potentials are generated by relativistic effects (\lq frame dragging\rq ) close to the neutron-star surface 
(Muslimov and Tsygan, 1990, 1992). As a consequence of the described electrodynamics, high-energy particles (Lorentz 
factors $\sim 10^6 - 10^7$ in most theories, $\sim 10^3$ in some) are accelerated in the gap regions and propagate 
along the magnetic field. Radiation losses due to magnetic bremsstrahlung (synchro-curvature radiation) and inverse 
Compton scattering on a low energy photon environment counterbalance the acceleration. If the energies in the 
particle/photon cascade exceed the threshold for pair creation ($\gamma - \gamma$  and $\gamma-\vec{B} \rightarrow e^+ e^-$), the 
extent of the acceleration region is further limited by the formation of a dense electron--positron plasma (pair 
formation front) which screens the open potentials. The high-energy cascade of particles and photons propagates 
through the magnetosphere until the photons decouple from the particles (\lq pulsar photosphere\rq ) and reach
 infinity, i.e. our telescopes. The observation of pulsed high-energy radiation allows us therefore to look directly 
at the particle spectra and magnetic field geometries on the \lq pulsar photosphere\rq . This high-energy 
photosphere can be visualized as a complex three-dimensional radiating surface with high directivity. Its properties 
should depend critically on 
pulsar parameters like field strength, inclination of the magnetic axis, aspect angle of observer, etc. and of 
course on the model assumptions. Although the scenario of magnetospheric high-energy cascades in the open zones is 
generally accepted, a fully self-consistent, quantitative theory has not yet been achieved. The terminal energies 
and spectra reached by the radiating particles depends on many uncertain assumptions and the debate on the relative 
importance of \lq polar cap\rq \ and/or \lq outer gap\rq \ emission continues, with occasional suggestions that the 
pulsar wind zone could be an additional source of \grs. A good set of observational facts, including multi-wavelength 
spectra and lightcurves, will certainly help to clarify our understanding of these fascinating high-energy machines. 

In the following, we qualitatively scale the high-energy radiation output with the  current of relativistic 
particles flowing from the polar cap (\lq PC\rq ). The Goldreich--Julian current, formed by $n_{GJ}$ particles 
flowing with $c$ from the PC area $A_{PC} \sim \pi r^2_{NS} ({r_{NS}\over r_c})$ is given by   
\begin{equation}  
\dot N_{GJ} = A_{PC}\cdot n_{GJ}\cdot c \sim 1.7\times 10^{38} {\dot P}^{1\over 2} P^{-{3\over 2}}\  {\rm s}^{-1}. 
\end{equation} 
If this particle flow consists mainly of electrons with Lorentz factor $\sim 10^6$ the outflowing power is of the 
order of $\sim 10^{38} {\dot P}^{1\over 2} P^{-{3\over 2}}$ erg s$^{-1}$. It should be noted that the product ${\dot 
P}^{1\over 2} P^{-{3\over 2}}$ is proportional to $\sqrt{\dot E_{rot}}$ and also to the total voltage available over 
the pulsar's open field lines $\Phi$. 

We can now estimate the energy flux observable from a pulsar at distance $d$:
\begin{equation}  
{F_E = (\eta \cdot {\dot E_{rot}}^{\alpha})} /{4\pi f_{\Omega} d^2}
\end{equation}
where $\eta$ is an efficiency factor that describes the conversion of the energy input (rotational energy loss or GJ 
current) into $\gamma$-radiation, $\alpha$ an exponent for the energy source (=0.5 if $\dot N_{GJ}$ dominates, =1 if 
$\dot E_{rot}$ relates directly to the \gr luminosity), and $f_{\Omega}$ describes the solid angle fraction 
illuminated by the pulsar beam (a canonical value of $f_{\Omega}= 1/4\pi$, i.e. beaming into 1 sr is assumed). 
Figure 1 shows the distribution of expected fluxes from a current list of radio pulsars with different assumptions 
for their luminosities. Rough sensitivity limits for past (COS-B, EGRET) and future (GLAST) telescopes are also 
shown in fig. 1. COS-B with 4-5 pulsars and EGRET with about a dozen detections are at the expected level. For GLAST 
we would estimate between 30 and 100 new radio pulsar detections. 
                
\begin{figure}
\centerline{\psfig{file=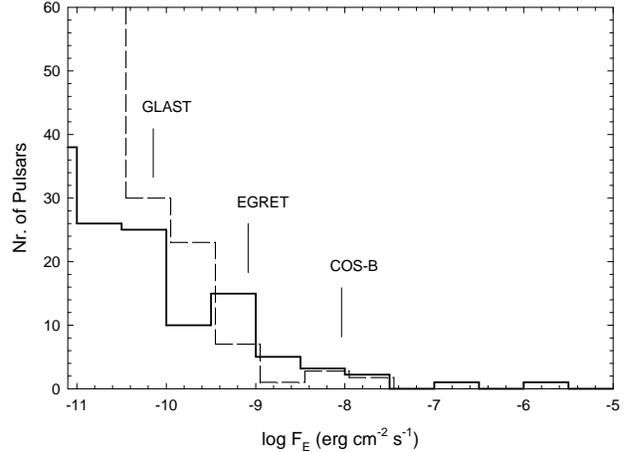,width=8.5cm,clip=} }
\caption{Distribution of $F_E$ for radio pulsars. For the $\gamma$-ray efficiency $\eta$ two assumptions are used:  
$\eta = \mbox{const} = 0.01$({\it solid line}) and $\eta{\dot E_{rot}}^{\alpha} \approx \sqrt{\dot E_{rot}}$ ({\it dashed 
line}). Approximate sensitivity limits for COS-B, EGRET and GLAST are shown}
\end{figure}

These predictions for high-energy emission from radio pulsars were published early after their discovery and the 
search for pulsars outside the radio range began. Detections were first reported for the Crab pulsar at X-ray (from 
balloon-borne detectors) and at optical wavelengths. Today more than 30 pulsars are known at X-ray energies. In the 
optical range, four to five pulsating counterparts and a similar number of optical candidates based on positional 
coincidence have been found.  

\begin{table*}[htb]

\begin{center}
\caption [Detections of high-energy pulsars]
{\label{tab01} {High-energy pulsars: multi-wavelength detections,
candidates, and positional coincidences \\
 Notes: P: pulsed emission; D: positional coincidence; ?: low significance pulsed detection}}
\begin{tabular}[htb]{l@{\hspace{3mm}}rrc@{\hspace{3mm}}c@{\hspace{3mm}}c@{\hspace{3mm}}c@{\hspace{3mm}}c@
{\hspace{3mm}}c@{\hspace {3mm}}c}
\hline
PSR & \multicolumn{1}{c}{$P$} & $\dot E / d^2$  & Radio & Optical & EUV & X$_{low}$ & X$_{hi}$ & $\gamma_{low}$ & $\gamma$ 
$_{hi}$ \\
 & (ms) &  rank & & & &  &  & &  \\
\hline
 & \multicolumn{9}{l}{High-confidence \gr detections} \\
\hline
B0531+21 (Crab) &  33.4 & 1 & P  & P & & P & P & P  & P$^1$ \\
B0833-45 (Vela) &  89.3 & 2 & P  & P & & P & & P  & P$^3$ \\
J0633+1746 (Geminga)& 237.1 & 3 & ?& ? & P & P & & ?$^9$  & P$^7$ \\
B1706-44  & 102.5 & 4 & P  & & & P & &  & P$^4$ \\
B1509-58  & 150.7 & 5 & P  & D & & P & P & P$^2$  & \\
B1951+32  &  39.5 & 6 & P  & & & P & & P$^{10}$ & P$^5$ \\
B1055-52  & 197.1 &33 & P  & D & & P & & P$^{11}$ & P$^8$ \\
\hline
 & \multicolumn{9}{l}{Candidate \gr detections} \\
\hline
B0656+14  & 384.9 &18 & P & ?& D & P & & ?$^{10}$ & ?$^6$ \\
B0355+54  & 156.4 &36 & P &   &   & D & &          & ?$^{12}$ \\
B0631+10  & 287.7 &53 & P &   &   &   & &          & ?$^{13}$ \\
B0144+59  & 196.3 &120& P &   &   &   & &          & ?$^{14}$ \\
\hline
 & \multicolumn{9}{l}{Candidate ms-pulsars \gr detections} \\
\hline
J0218+4232& 2.32  &43 & P &   &   & P & &          & ?$^{15}$ \\
B1821-24  & 3.05  &14 & P &   &   & P & &          & ?$^{16}$ \\
\hline
 & \multicolumn{9}{l}{Likely pulsar--\gr source coincidences} \\
\hline
B1046-58  & 123.7  &8 & P &   &   & D & &          & ?$^{18}$ \\
J1105-6107&  63.2  &21 & P$^{17}$ &   &   & D & &          &  \\
B1853+01   & 267.4  &27 & P &   &   &  ? & &          & ? \\
\hline
\multicolumn{1}{l}{References:} \\
 \multicolumn{10}{l}{$^1$ Nolan et al.(1993); $^2$ Matz et al.(1994), Kuiper et al.(1999a); $^3$  Kanbach et 
al.(1994);} \\
 \multicolumn{10}{l}{$^4$ Thompson et al.(1992, 1996); $^5$ Ramanamurthy et al.(1995); $^6$ Ramanamurthy et 
al.(1996);} \\
 \multicolumn{10}{l}{$^7$ Mayer-Hasselwander et al.(1994); $^8$ Fierro et al.(1993); $^9$ Kuiper et al.(1996);} \\
 \multicolumn{10}{l}{  $^{10}$ Hermsen et al.(1997); $^{11}$ Thompson et al.(1999); $^{12}$ Thompson et al.(1994);} 
\\
 \multicolumn{10}{l}{$^{13}$ Zepka et al.(1996); $^{14}$ Ulmer et al.(1996); $^{15}$ Kuiper et al.(1999b);} \\
 \multicolumn{10}{l}{ $^{16}$ Thompson et al.(1997); $^{17}$ Kaspi et al.(1997); $^{18}$ Kaspi et al.(2000)} \\
\end{tabular}
\end{center}
\end{table*}

\section{Detections}

Table 1 (Kanbach, 2001) gives an overview of the high-energy pulsars and their multi-frequency detections, ranging 
from the radio band to gamma rays, after the decade of discoveries with the CGRO instruments. 
Next to the column with the spin period the rank of the quantity $\dot E / d^2$, i.e. the spin-down power over the 
square of the distance as a measure of the apparent brightness, is given. It is evident that the detections in the 
gamma-ray range follow closely the ranking derived from the spin-down luminosity. Therefore we might conclude that 
gamma-ray emission is a common phenomenon of pulsars. However, since so far most detections of gamma-ray pulsars 
require the detection in the radio or X-ray regime in order to provide the periodicity ephemeris, we must be 
careful not to discard the possibility that many pulsars might exist that emit dominantly at high energies and so 
far have only been detected as unidentified \gr sources. 

\section{Emission Characteristics}
\subsection{Lightcurves}
The regular  brightness variation of pulsars ('lightcurves'), with periods ranging from $\sim ms$ to nearly 10s, was 
the primary characteristic of this new astrophysical object.   
The lightcurve of a pulsar can be regarded as the cross-section through a rotating beaming pattern cut along the  
line of sight to the observer. The full beaming pattern on the sphere around a pulsar would give us a direct view of 
the sites of origin and beaming directions and allow us to derive the luminosity of such a 'highly' non-isotropic 
source. But of course we are restricted to our singular view of individual sources. In order to gain a better 
understanding of high energy pulsars we can follow two routes: (1) rely on theoretical modeling or (2) observe a 
sufficient number of pulsars to get a statistical relevant sample with different aspect angles. The rather limited 
number of detections has not yet allowed a full population treatment. But some common traits for the \gr emission 
from young pulsars  become already apparent. Figure 2 displays the high-energy lightcurves of the brightest 6 EGRET 
pulsars. We note that the pulsed emission extends over more than 50\%  of rotation and sharp peaks (generally 2) 
limit the 'pulsed' phase interval. The width and dominance of the peaks led to the conventional assumption of a 
beaming solid angle of 1 sr for luminosity estimates. The likely beaming pattern inferred from these \gr lightcurves 
is in the shape of a cone with high intensity on the rim and lower levels inside ('hollow cone'). Both leading 
models, polar cap and outer gap, generate this kind of pattern, although the 2-dimensional shape for the outer gap 
model could be far from circular. The bright rims of the emission cones indicate the location of intense 
particle-photon cascades in the magnetospheres. The emission inside the cones is often found to have a harder 
spectrum, which could indicate that this radiation has been propagated along magnetic field lines with less 
curvature (polar cap region) avoiding the cascading processes. 
               
\begin{figure}
\psfig{file=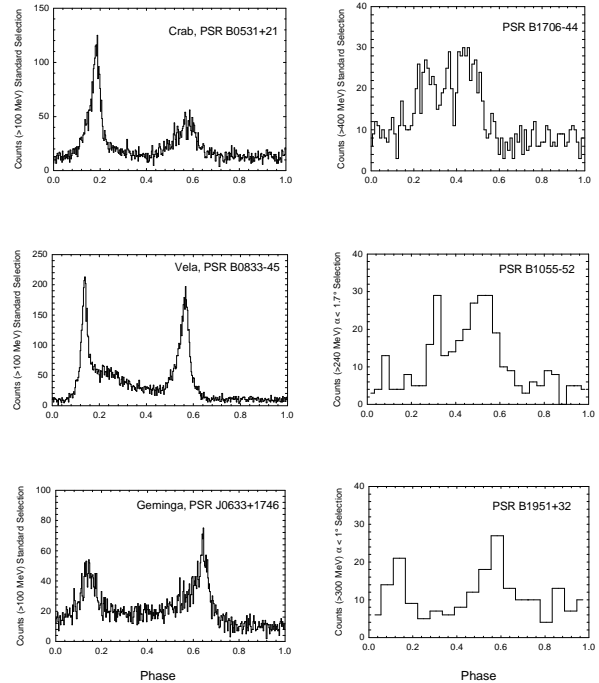,width=8.5cm,clip=}
\caption{High-energy light curves of $\gamma$-ray pulsars ($>100$
MeV, unless indicated differently)
\label{image}}
\end{figure}
 
In the following we compare the \gr lightcurves of the three brightest \gr  pulsars  with their emission at lower 
energies. These multi-wavelength comparisons serve to distinguish different emission processes. Figure 3 (right 
panel) shows the Crab pulsar between optical and \gr energies. The overall shape of the double-peaked lightcurve 
remains similar over this range but spectral differences between the components are clearly visible: the bridge 
region between the two peaks and the second peak are very strong between 100 keV and $\sim $10 MeV. Also in the 
radio range (left panel figure 3) we find additional structures besides the two peaks: two trailing peaks around 5 
GHz and a so-called precursor peak at frequencies below 400 MHz. No comprehensive model has yet been proposed that 
can explain this complex spatial and spectral emission but it is apparent that several components must exist in the 
radiation zones of the Crab pulsar.     
      
\begin{figure}
\psfig{file=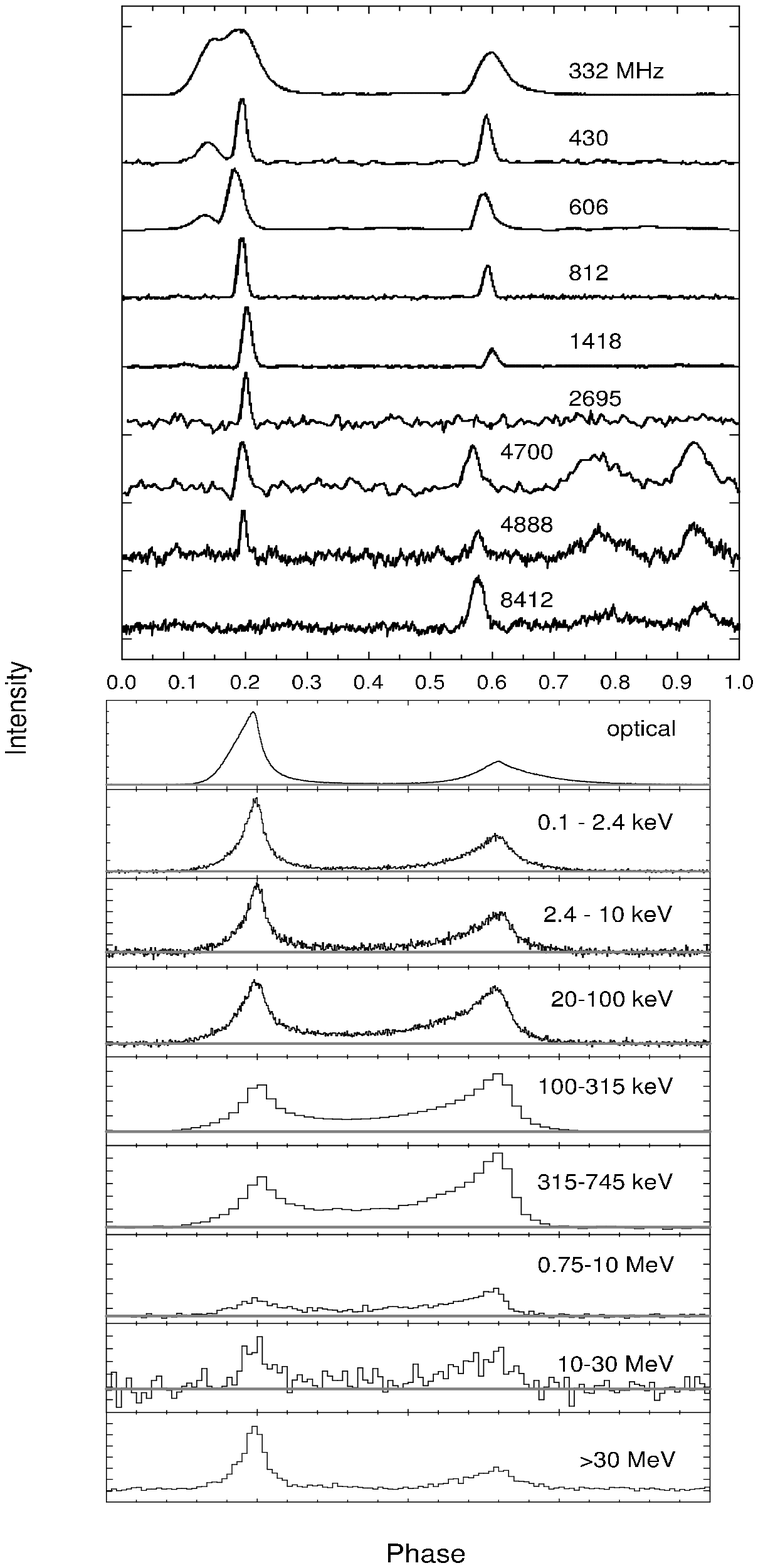,width=8.5cm,clip=} 
\caption{Multi-wavelength lightcurves of the Crab pulsar in the radio range (top, Moffett \&  Hankins, 1996) and at 
high energies (bottom, Kuiper et al., 2001).
\label{image}}
\end{figure}

The same conclusions can be derived for the older pulsars Vela and Geminga. Their multi-wavelength lightcurves are 
shown in figure 4. Vela is clearly detected as a pulsating source from radio to \gr energies. The lightcurves below 
the $\gamma$ range are however, in contrast to the Crab case, completely at different phase angles of emission. The same 
finding applies to the Geminga pulsar. In this case significant pulsed emission has only been detected in the X- and 
\gr region. The present optical and radio detections of Geminga are considered to be of marginal significance and 
need to be confirmed. We conclude again, that the available multi-wavelength lightcurves show that the 
magnetospheres of young pulsars contain several additional emitting regions with different spectral and beaming 
characteristics on top of a basic two-peaked structure. 

\begin{figure}
\psfig{file=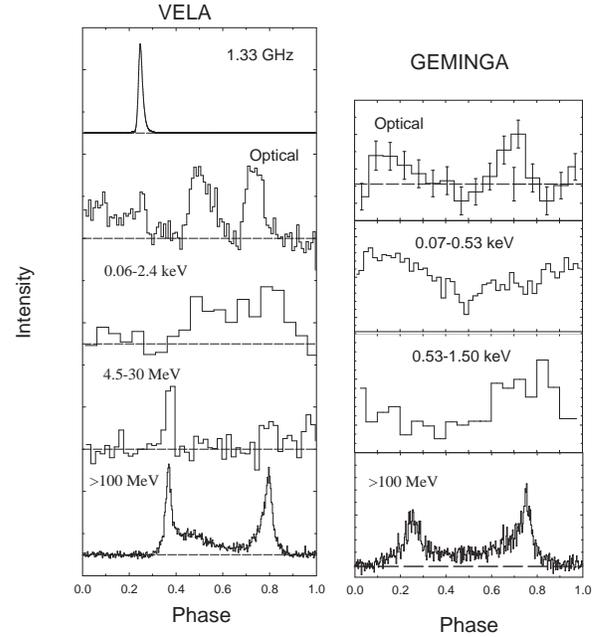,width=8.5cm,clip=} 
\caption{Multi-wavelength lightcurves of the Vela and Geminga pulsars
\label{image}}
\end{figure}

\subsection{Spectra}
Thompson et al., 1999 have compiled multi-wavelength spectra for the pulsed emission of \grpsrs as shown in figure 5. 
The spectral intensities are displayed in terms of $\nu F_{\nu }$ , which is equivalent to the power emitted per 
unit interval of $ln(\nu)$. The spectra emphasize that emission in the X- and \gr regions dominates the radiation 
budget of these pulsars and that the pulsed emissions are generally harder than power-laws with index -2 (which 
would be a horizontal line in this spectral display). In several cases the spectra are curved in the \gr range and 
lead up to a cut-off or turn over at a few GeV. The pulsar spectra are arranged in order of increasing rotational 
age ($\tau \sim {P\over {\dot P}}$) with Crab ($\tau \sim 10^3 y$) as the youngest and PSR B1055-52 ($\tau \sim 
5\times 10^5 y$) as the oldest object.

The {\bf Crab} pulsar  emits the maximum power at about 100 keV and the spectrum can be described by a broad peak 
extending from the optical band ($\approx 1$ eV) to about 100 MeV. Above 100 MeV the pulsed spectrum continues with 
a harder power law distribution up to a few GeV, where the spectrum goes into a steep decline (Kuiper et al., 2001). 
In addition to the pulsed emission, the Crab shows a strong unpulsed component at high energies. This component has 
been interpreted to come from the inner Crab nebula as synchrotron radiation below a few GeV and as inverse Compton 
radiation, up-scattered from optical photons, up to TeV energies.

{\bf PSR B1509-58} has been detected in the hard X-ray range by Ginga and at low \gr energies up to 30 MeV by BATSE, 
OSSE and COMPTEL. EGRET may have a marginal detection of this pulsar between 30-70 MeV  but places significant upper 
limits on the flux above this energy range, which indicates that this spectrum must turn over in the $\sim 10$ MeV 
range. This low energy spectral cut-off in PSR B1509-58 has been explained as an absorption effect for high-energy 
photons in the magnetic field. PSR B1509-58 has one of the strongest inferred magnetic fields ($\sim 2\times 
10^{13}$ Gauss, about 40\% of $B_{crit}$) and could be called a 'magnetar'. Harding et al., 1997 have modelled polar 
cap photon cascades in such strong fields and find that the second order quantum-mechanical process of 'photon 
splitting' ($\gamma \rightarrow \gamma \gamma$) would limit the energy of emitted \grs to several 10's of MeV.

The {\bf Vela} pulsar (PSR B0833-45) has a pronounced spectral break at 2 GeV. Due to the strength of this source, 
very detailed spectra for the individual phase components, e.g. the peaks and the interpulse emission, are 
available. The spectra of the emission peaks are generally softer than the spectra of the inter-peak regions. The 
softest components are observed in the leading and trailing wings of the peaks, which has been explained as a 
low-energy spill-out from the main \gr cascades in the outer magnetosphere.

{\bf PSR B1706-44}, a \gr source that had been detected with COS-B 25 years ago (2CG342-02), was identified in 1992 
as a radio pulsar with a 102 ms pulsation period. Although an X-ray source is found coincident with the pulsar no 
pulsation is found at keV energies. Also at TeV energies the source is detected only as a steady emitter, similar to 
the Crab. The maximum power of PSR1706-44 is emitted around 1 GeV.

{\bf PSR B1951+32} is found as a weak \gr source at energies above 300 MeV. After extended EGRET observations, 
pulsational analysis was successful and confirmed the identification of this \gr source with the radio pulsar. The 
spectrum of this pulsar shows the characteristic spectral break above several GeV only marginally, because of the 
limited statistics for this source .

{\bf Geminga} (PSR B0633+17) was the most enigmatic \gr source since its detection with SAS-2 in 1972. After long 
and patient searches with X-ray and optical telescopes Geminga was finally identified as a pulsar in 1992. ROSAT 
observations revealed a 237 ms pulsations in Geminga (Halpern and Holt, 1972) and the corresponding \gr signal was 
found in data from EGRET, COS- B, and SAS-2. The spectrum of the pulsed emission is generally very hard and shows 
marked variations over the rotation period. The maximum power of Geminga is emitted at about 1 GeV. Above a few GeV 
the spectrum breaks off sharply. Geminga is the first specimen of a true high-energy pulsar. The power in optical 
($m_v\sim 25.5$, pulsation marginally detected) and radio emissions (marginal detections of pulsed emission at 102 
MHz) is lower than the \gr emission by more than 6 orders of magnitude, justifying the designation 'radio-quiet 
pulsar' for Geminga. Speculations that many of the other unidentified galactic \gr sources, which appear with 
comparable brightness, are similar \lq Geminga-like\rq \ pulsars have been discussed widely. One must realize 
however that Geminga is a very close ($\sim 160$ pc), low-luminosity pulsar and that the other \gr sources, based on 
their galactic distribution are at least 10 times more distant and therefore should be a hundred times more luminous 
than Geminga. If the above hypothesis is maintained one has to assume that much younger, high-luminosity pulsars can 
also operate in a \lq Geminga-like\rq \  mode. The absence of significant radio- and X-ray counterpart detections 
for most of the unidentified \gr sources requires also that the emission patterns at different wavelengths would not 
coincide with the $\gamma$ beams. 

{\bf PSR B1055-52} also has a very hard energy spectrum that seems to extend from  X-rays into the \gr range. Again 
the maximum power is emitted around 1 GeV. No clear break in the spectrum is visible up to 4 GeV. Above that energy 
a break is required by upper limits at TeV energies.

Figure 6, with an enlarged version of the pulsed spectrum from the Crab, shows the current concept of spectral 
components in high-energy pulsars. The low energy spectrum, with a peak around 100 keV, is thought to result from 
synchrotron emission. Based on the spectral maximum of emission from a relativistic electron ($E=\gamma  mc^2$) in a 
magnetic field $B_\bot$ which is at $\nu _{max} \sim 1.2\  MHz\cdot B_\bot (G)\cdot \gamma ^2$ an order of magnitude 
estimate for the peak at 100 keV demands that the product $B_\bot (G)\cdot \gamma ^2$ is about $2\times 10^{13}$. 
Such values could be achieved either with very energetic electrons ($\gamma \sim 10^{6-7}$) in outer magnetospheric 
fields of $\sim 1$ Gauss or with electrons of ($\gamma \sim 10^{2-3}$) in inner polar fields with  $B_\bot \sim 
10^8$ Gauss.
The high energy peak could result from inverse Compton scattering of energetic electrons on low energy radiation of 
energy $\varepsilon$. The typical energy of the inverse Compton photons is then $E_{\gamma} ={4\over 3}\epsilon 
\gamma ^2$, which means that electrons with Lorentz factors $\sim 10^4$ could boost 10 eV photons into the GeV 
range.

As we saw from the multi-wavelength lightcurves pulsars show spectral changes as they rotate, i.e. the lightcurves 
look different for adjacent energy bands. This phase dependent spectroscopy can be extended to the \gr range also. 
In figure 7 a hardness ratio, defined as the ratio between the number of photons above 300 MeV to those in the band 
100-300 MeV, is shown as a function of the pulsar's rotation. We notice that in the prominent '2-peak' pulsars the 
first peak has a softer spectrum than the interval region. The second peak is generally also harder than the first 
peak. Model calculations for polar cap cascades (Daugherty and Harding, 1996) as well as the outer-gap model of 
Romani, 1996 have successfully reproduced these phase-dependent spectra for Vela. 

\begin{figure}
\psfig{file=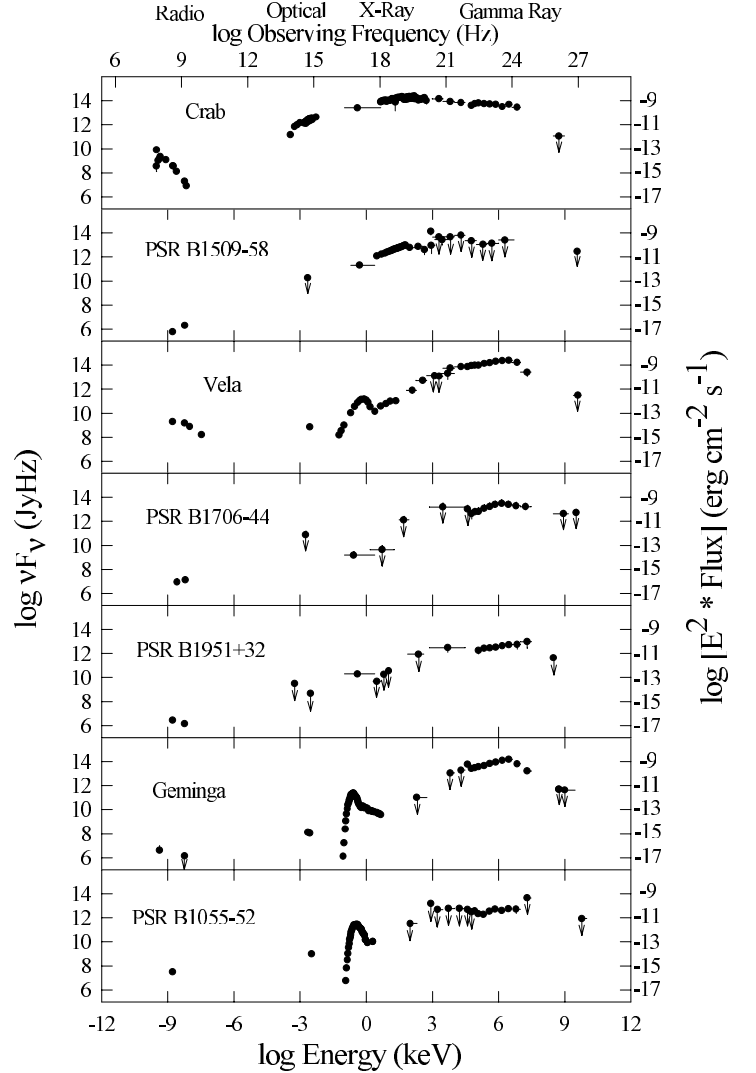,width=9.5cm,clip=} 
\caption{Multi-wavelength spectra (pulsed emission only) of $\gamma$-ray pulsars (Thompson et al., 1999). 
\label{image}}
\end{figure}

\begin{figure}
\psfig{file=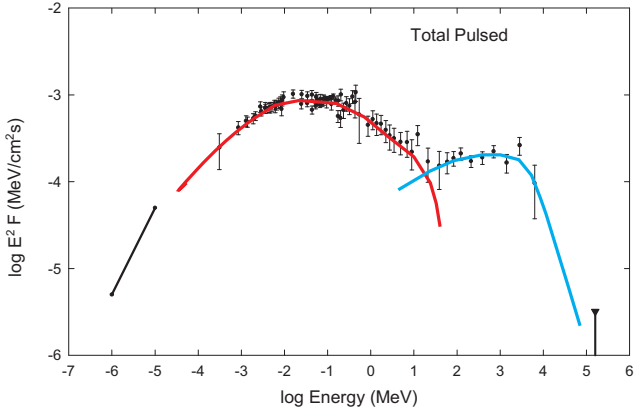,width=8.5cm,clip=} 
\caption{Pulsed emission from the Crab pulsar with theoretical spectra corresponding to emission via the synchrotron 
process (low energy peak) and the inverse Compton or curvature radiation peak at higher energies ( data from Kuiper 
et al., 2001). 
\label{image}}
\end{figure}

\begin{figure}
\psfig{file=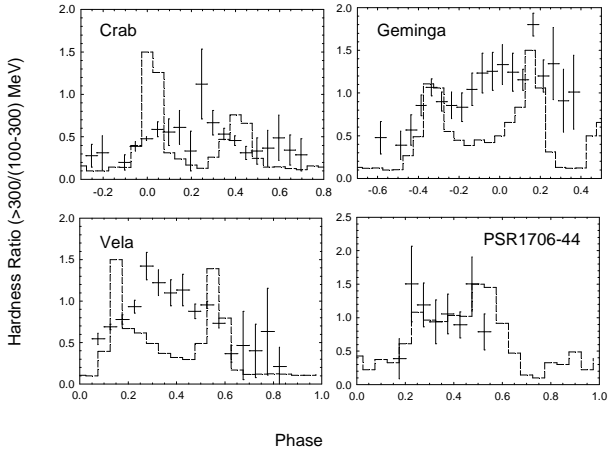,width=8.5cm,clip=} 
\caption{Spectral hardness as a function of rotation phase for 4 bright \grpsrs. The location of the emission peaks 
is shown in the underlying lightcurves.
\label{image}}
\end{figure}

\subsection{Luminosities and Energetics}

The energy budget of high-energy of pulsars is derived from the pulsed emission spectra by integration from optical 
wavelengths ($\sim $1 eV) to the cut-off at GeV energies. The result is given in table 2 where $F_E$ is the observed 
high-energy flux. For the calculation of the efficiency $\eta $, with which the pulsar generates energetic photons 
from the loss of rotational energy, we assume that the emission is beamed into a solid angle of 1 sr. This could be 
a problematic assumption especially for the wide beaming patterns generated in outer gap models. Figure 8 shows 
plots of the correlation of the high-energy properties, spectral index and efficiency ($\geq 100$ MeV) with some 
derived pulsar parameters: the rotational age ($\tau\propto P/\dot P$), the magnetic field ($B\propto\sqrt{P \dot 
P}$) and the total voltage available over the pulsar's open field lines $( \Phi\propto \sqrt{\dot P / P^3})$. As 
many previous investigations have shown, both the efficiencies and the spectral indices for radiation above 100 MeV 
correlate well with the apparent age and the open field line potentials of pulsars. Pulsars become more efficient in 
their conversion of rotational energy into $E>100$ MeV radiation and their spectra become harder with increasing age 
or decreasing  potential values. The correlation with the inferred magnetic field is not as clear. Indeed, there 
seems to be an indication of a decreasing efficiency for both the highest $B$ fields and for  low $B$ values. 
Similarly the spectra appear softer at the marginal magnetic field values. Figure 9 shows directly the correlation 
between the total high-energy luminosity ($\geq $1 eV) and the Goldreich-Julian current. The direct proportionality 
of $L_{\gamma} \propto \dot E^{0.5}$ is in contrast to the relation at soft X-ray energies $L_X \propto \dot E$ 
(Becker and Tr\" umper, 1997), which could indicate different emission mechanisms in the two bands. As we see in 
figure 8, the efficiency of some older pulsars appears already quite high ($\sim 20\%$) and a limit to
 the conversion of rotational energy into \grs must be reached at some value of the open field line potential ($\Phi 
\leq 10^{14}$ V). Such a condition could indicate the termination of \gr production in old pulsars and form a 
so-called 'death-line' in the conventional $P-\dot P$ diagram. Only future, more sensitive, \gr observations will be 
able to define this transition from a '\gr -active' to a '\gr -dead' pulsar more clearly.

\begin{table*}
\caption[Energetics of high-energy pulsars]
{\label{tab02} 
Properties of high-energy pulsars (Thompson et al., 1999)}
\begin{tabular}{rrrlllll}
\cline{1-8}
\cline{1-8}
\vbox to1.70ex{\vspace{1pt}\vfil\hbox to8.0ex{ Name\hfil}} &
\vbox to1.70ex{\vspace{1pt}\vfil\hbox to6.4ex{\hfil $P$\hfil}} &
\vbox to1.70ex{\vspace{1pt}\vfil\hbox to8.0ex{\hfil $\tau$\hfil}} &
\vbox to1.70ex{\vspace{1pt}\vfil\hbox to11.0ex{\hfil $\dot E$\hfil}} &
\vbox to1.70ex{\vspace{1pt}\vfil\hbox to13.0ex{\hfil $F_E$\hfil}} &
\vbox to1.70ex{\vspace{1pt}\vfil\hbox to6.0ex{\hfil $d$\hfil}} &
\vbox to1.70ex{\vspace{1pt}\vfil\hbox to10.0ex{\hfil $L_{HE}$\hfil}} &
\vbox to1.70ex{\vspace{1pt}\vfil\hbox to8.0ex{\hfil $\eta$\hfil}} \\


\vbox to1.70ex{\vspace{1pt}\vfil\hbox to8.0ex{\hfil \hfil}} &
\vbox to1.70ex{\vspace{1pt}\vfil\hbox to6.4ex{\hfil (ms)}} &
\vbox to1.70ex{\vspace{1pt}\vfil\hbox to8.0ex{\hfil ($10^3$ yr)\hfil}} &
\vbox to1.70ex{\vspace{1pt}\vfil\hbox to11.0ex{\hfil ($10^{36}$ erg s$^{-1}$)\hfil}} &
\vbox to1.70ex{\vspace{1pt}\vfil\hbox to13.0ex{\hfil (erg cm$^{-2}$ s$^{-1}$ )\hfil}} &
\vbox to1.70ex{\vspace{1pt}\vfil\hbox to6.0ex{\hfil (kpc)\hfil}} &
\vbox to1.70ex{\vspace{1pt}\vfil\hbox to10.0ex{\hfil (erg s$^{-1}$)\hfil}} &
\vbox to1.70ex{\vspace{1pt}\vfil\hbox to8.0ex{\hfil ($E>1$eV)\hfil}\vspace{1pt}} \\

\cline{1-8}
\vbox to1.70ex{\vspace{1pt}\vfil\hbox to8.0ex{ Crab\hfil}} &
\vbox to1.70ex{\vspace{1pt}\vfil\hbox to6.4ex{\hfil 33}} &
\vbox to1.70ex{\vspace{1pt}\vfil\hbox to9.0ex{\hfil 1.3\hfil}} &
\vbox to1.70ex{\vspace{1pt}\vfil\hbox to10.0ex{450}} &
\vbox to1.70ex{\vspace{1pt}\vfil\hbox to13.0ex{\hfil $1.3\times 10^{-8}$\hfil}} &
\vbox to1.70ex{\vspace{1pt}\vfil\hbox to6.0ex{\hfil 2.0\hfil}} &
\vbox to1.70ex{\vspace{1pt}\vfil\hbox to10.0ex{\hfil $5.0\times 10^{35}$\hfil}} &
\vbox to1.70ex{\vspace{1pt}\vfil\hbox to8.0ex{\hfil 0.001\hfil}} \\

\vbox to1.70ex{\vspace{1pt}\vfil\hbox to8.0ex{ B1509-58\hfil}} &
\vbox to1.70ex{\vspace{1pt}\vfil\hbox to6.4ex{\hfil 150}} &
\vbox to1.70ex{\vspace{1pt}\vfil\hbox to9.0ex{\hfil 1.5\hfil}} &
\vbox to1.70ex{\vspace{1pt}\vfil\hbox to10.0ex{18}} &
\vbox to1.70ex{\vspace{1pt}\vfil\hbox to13.0ex{\hfil $8.8\times 10^{-10}$\hfil}} &
\vbox to1.70ex{\vspace{1pt}\vfil\hbox to6.0ex{\hfil 4.4\hfil}} &
\vbox to1.70ex{\vspace{1pt}\vfil\hbox to10.0ex{\hfil $1.6\times 10^{35}$\hfil}} &
\vbox to1.70ex{\vspace{1pt}\vfil\hbox to8.0ex{\hfil 0.009\hfil}} \\

\vbox to1.70ex{\vspace{1pt}\vfil\hbox to8.0ex{ Vela\hfil}} &
\vbox to1.70ex{\vspace{1pt}\vfil\hbox to6.4ex{\hfil 89}} &
\vbox to1.70ex{\vspace{1pt}\vfil\hbox to8.0ex{\hfil 11\hfil }} &
\vbox to1.70ex{\vspace{1pt}\vfil\hbox to11.0ex{7.0}} &
\vbox to1.70ex{\vspace{1pt}\vfil\hbox to13.0ex{\hfil $9.9\times 10^{-9}$\hfil}} &
\vbox to1.70ex{\vspace{1pt}\vfil\hbox to6.0ex{\hfil 0.5\hfil}} &
\vbox to1.70ex{\vspace{1pt}\vfil\hbox to10.0ex{\hfil $2.4\times 10^{34}$\hfil}} &
\vbox to1.70ex{\vspace{1pt}\vfil\hbox to8.0ex{\hfil 0.003\hfil}} \\

\vbox to1.70ex{\vspace{1pt}\vfil\hbox to8.0ex{ B1706-44\hfil}} &
\vbox to1.70ex{\vspace{1pt}\vfil\hbox to6.4ex{\hfil 102}} &
\vbox to1.70ex{\vspace{1pt}\vfil\hbox to8.0ex{\hfil 17\hfil }} &
\vbox to1.70ex{\vspace{1pt}\vfil\hbox to11.0ex{3.4}} &
\vbox to1.70ex{\vspace{1pt}\vfil\hbox to13.0ex{\hfil $1.3\times 10^{-9}$\hfil}} &
\vbox to1.70ex{\vspace{1pt}\vfil\hbox to6.0ex{\hfil 2.4\hfil}} &
\vbox to1.70ex{\vspace{1pt}\vfil\hbox to10.0ex{\hfil $6.9\times 10^{34}$\hfil}} &
\vbox to1.70ex{\vspace{1pt}\vfil\hbox to8.0ex{\hfil 0.020\hfil}} \\

\vbox to1.70ex{\vspace{1pt}\vfil\hbox to8.0ex{ B1951+32\hfil}} &
\vbox to1.70ex{\vspace{1pt}\vfil\hbox to6.4ex{\hfil 40}} &
\vbox to1.70ex{\vspace{1pt}\vfil\hbox to8.0ex{\hfil 110\hfil }} &
\vbox to1.70ex{\vspace{1pt}\vfil\hbox to11.0ex{3.7}} &
\vbox to1.70ex{\vspace{1pt}\vfil\hbox to13.0ex{\hfil $4.3\times 10^{-10}$\hfil}} &
\vbox to1.70ex{\vspace{1pt}\vfil\hbox to6.0ex{\hfil 2.5\hfil}} &
\vbox to1.70ex{\vspace{1pt}\vfil\hbox to10.0ex{\hfil $2.5\times 10^{34}$\hfil}} &
\vbox to1.70ex{\vspace{1pt}\vfil\hbox to8.0ex{\hfil 0.007\hfil}} \\

\vbox to1.70ex{\vspace{1pt}\vfil\hbox to8.0ex{ Geminga\hfil}} &
\vbox to1.70ex{\vspace{1pt}\vfil\hbox to6.4ex{\hfil 237}} &
\vbox to1.70ex{\vspace{1pt}\vfil\hbox to8.0ex{\hfil 340\hfil }} &
\vbox to1.70ex{\vspace{1pt}\vfil\hbox to11.0ex{0.033}} &
\vbox to1.70ex{\vspace{1pt}\vfil\hbox to13.0ex{\hfil $3.9\times 10^{-9}$\hfil}} &
\vbox to1.70ex{\vspace{1pt}\vfil\hbox to6.0ex{\hfil 0.16\hfil}} &
\vbox to1.70ex{\vspace{1pt}\vfil\hbox to10.0ex{\hfil $9.6\times 10^{32}$\hfil}} &
\vbox to1.70ex{\vspace{1pt}\vfil\hbox to8.0ex{\hfil 0.029\hfil}} \\

\vbox to1.70ex{\vspace{1pt}\vfil\hbox to8.0ex{ B1055-52\hfil}} &
\vbox to1.70ex{\vspace{1pt}\vfil\hbox to6.4ex{\hfil 197}} &
\vbox to1.70ex{\vspace{1pt}\vfil\hbox to8.0ex{\hfil 530\hfil }} &
\vbox to1.70ex{\vspace{1pt}\vfil\hbox to11.0ex{0.030}} &
\vbox to1.70ex{\vspace{1pt}\vfil\hbox to13.0ex{\hfil $2.9\times 10^{-10}$\hfil}} &
\vbox to1.70ex{\vspace{1pt}\vfil\hbox to6.0ex{\hfil 1.5\hfil}} &
\vbox to1.70ex{\vspace{1pt}\vfil\hbox to10.0ex{\hfil $6.2\times 10^{33}$\hfil}} &
\vbox to1.70ex{\vspace{1pt}\vfil\hbox to8.0ex{\hfil 0.207\hfil}\vspace{1pt}} \\

\cline{1-8}
\end{tabular}
\end{table*}

\begin{figure}
\psfig{file=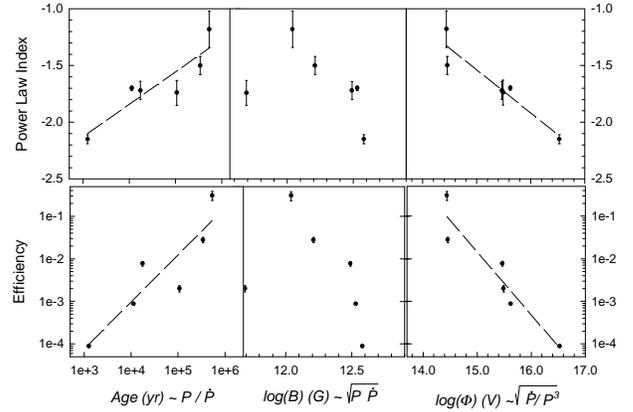,width=8.5cm,clip=} 
\caption{Correlation of several derived pulsar parameters (rotational age, magnetic field strength, and potential 
across the polar cap, which is also proportional to the Goldreich-Julian current) with the observed intensity and 
hardness of $\gamma$ radiation above $\sim 100 MeV$ .
\label{image}}
\end{figure}

\begin{figure}
\psfig{file=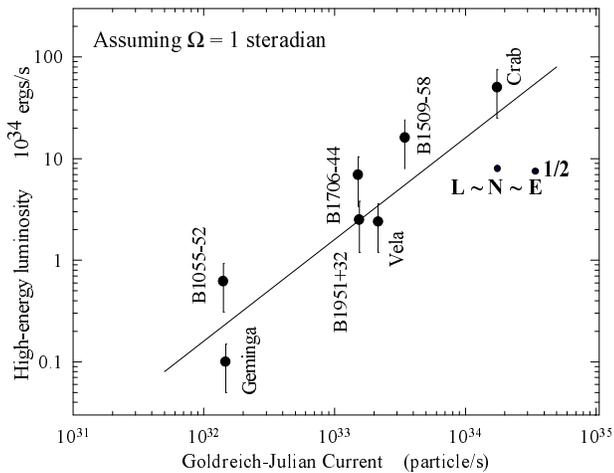,width=8.5cm,clip=} 
\caption{Pulsar luminosity above $\sim 1$ eV, assuming a beaming pattern solid angle
of 1 sr, versus the Goldreich--Julian current flowing from the open magnetosphere.
(Thompson et al., 1999).
\label{image}}
\end{figure}

\section{Conclusions and Outlook }

After the mission of the Compton Gamma Ray Observatory we know about a dozen (7 definitely and a few more as likely 
candidates) \gr pulsars. These are objects in which the most extreme electromagnetic and gravitational conditions in 
the universe act to accelerate highly relativistic particles, leading to high-energy luminosities of the order of 
$10^{35}$ erg/s. Young \grpsrs can in principle be observed throughout the Galaxy because \grs suffer neither 
absorption nor dispersion of the pulsed signals. High-energy radiation from magnetized rotating neutron stars is not 
only a fascinating astrophysical topic in itself -- a thorough understanding of the radiation processes also opens  
the way to uncover the true nature of some of the presently unidentified \gr sources. The majority of these objects 
are galactic (their distribution correlates well with the galactic disk) but the $\sim 170$ objects of the EGRET 
catalog ($\sim 60\%$ of all \gr sources) pose a major challenge and puzzle to astronomy. It is known that the \gr  
sources correlate generally with regions containing young populations like molecular clouds, star forming regions, 
OB associations, HII regions, or SNRs. We can safely assume that these regions also contain young pulsars that have 
not yet been revealed, either being hidden inside their birth places or for a lack of emission at other wavelengths. 
As described above, Geminga can be taken  as a r\^ole model for such a population of pure \gr pulsars. The discovery 
of individual young \grpsrs will certainly support our theoretical understanding of pulsar physics, but their 
population will also provide us with new insights into active star forming regions and the processes leading to the 
formation of neutron stars. Overall the contribution of unresolved, distant pulsars to the diffuse galactic \gr 
emission could be in the range of 5--30{\%}. 

The next generation of \gr telescopes, especially the GLAST project as a successor to EGRET (Michelson, 1996), will 
be much more sensitive and is expected to discover 30 - 100 new \grpsrs based on the catalog of radio pulsars alone.
With the much higher sensitivity of GLAST and the resulting high photon detection rate it will also be possible to 
investigate many unidentified \gr sources for periodicities independent of observations at other wavelengths. For 
EGRET this capability for independent periodicity detection has been demonstrated only for Geminga (Brazier and 
Kanbach, 1996; Chandler et al., 2001). None of the other EGRET sources has a  signal/noise ratio with a high photon 
detection rate similar to Geminga and the powerful investigation for periodicities in some strong EGRET sources 
carried out by Chandler et al., 2001 was unsuccessful. Thus the next generation of \gr telescopes has the potential 
to reveal the radiation physics of pulsars with a wide range of ages and to discover a new population of pure 
high-energy pulsars.


\clearpage

\end{document}